# SENSOR WEBS FOR ENVIRONMENTAL RESEARCH

Nishadh, K A – Azeez, P. A.

*Environmental Impact Assessment Division, Sálim Ali Center for Ornithology and Natural History (SACON), Anaikatty (PO), Coimbatore – 641108.*

*(phone: +91 0422-2203100; fax: +91 0422-2657088)*

e-mail: nishadhka@sacon.in

**Abstract.** The ongoing massive global environmental changes and the past learnings have highlighted the urgency and importance of further detailed understanding of the earth system and implementation of social ecological sustainability measures in a much more effective and transparent manner. This short communication discuss the potential of sensor webs in addressing those research challenges, highlighting it in the context of air pollution issues.
**Keywords:** *Sensor web, Earth system science, social-ecological system, Environmental monitoring*

**Introduction**

The idea of connecting hypertext digital documents/information over Internet and computer has realized the World Wide Web (Info.cern.ch 2008). Raising user experience from mere user of those documents to a provider and reviewer of documents/information has set forth the stage for next evolution in this line. This, named 'web 2.0' achieved expansion into unforeseen ventures in human history in terms of generating, sharing, editing, accessing, analyzing and re-synthesizing the information at high speed, in large scale and with dependable accuracy. Wikipedia, Flickr and Google maps are some of the resultant almost ubiquitous products of such ventures (Goodchild, 2007). Sensor webs stands as the next leap forward in that revolution gestated through merging Internet, World Wide Web and sensor system; a development from the human inscribed World Wide Web to sensor acquired data and information web about our immediate surroundings (Zyl, et al., 2009). It has great potential in better understanding the earth system science (Torres-Martinez, et al., 2003) and in ensuring social ecological system sustainability. It also would further better access to information, transparency and informed decision making.

**Sensors, Sensor Webs and Specifications:**

To better know about our environment, sensors are basic instrumental capacity to transform various types of environmental stimuli (physical, biological or chemical) into transmissible and interpretable signals. With ever increasing necessity and want for environmental information, numerous sensors, its aggregate system and networks are deployed for various purposes. Due to the inherent variability of sensors systems in terms of hardware specifications, protocols, interfaces and communication style, it become very difficult to utilize and manage the sensor resources to give holistic and dynamic environmental information (Bröring, et al., 2011; Havlik et al., 2011). The sensor web has solution for that (*Fig. 1*).

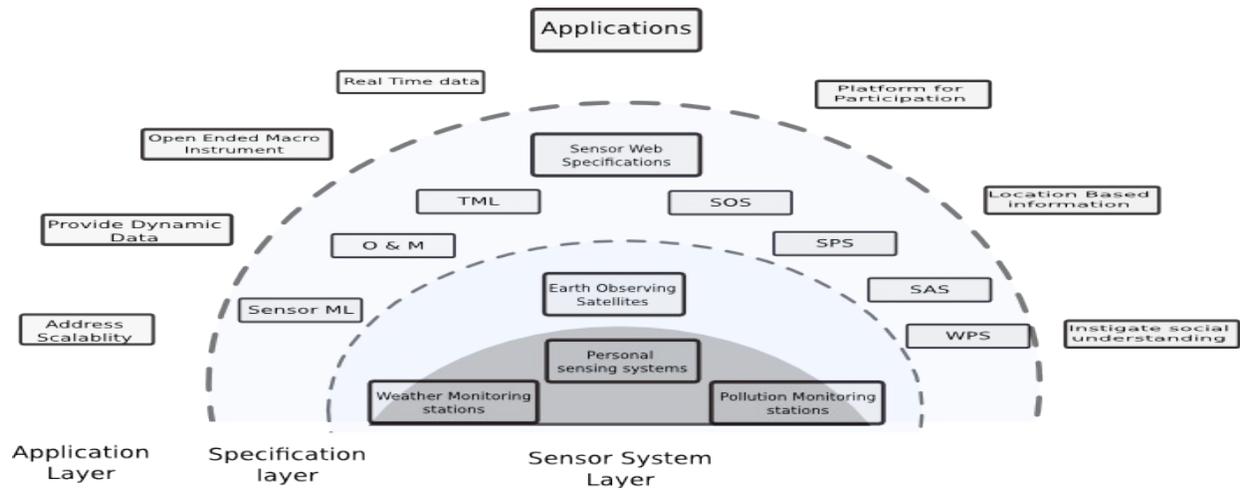

**Figure-1.** The different layers of sensor web enablement and its applications

As the name suggests, the sensor web is a web of sensors, sensor systems, its networks, data and metadata with capabilities for autonomous self-organization and adaptive capabilities to act as a coordinated macro-instrument(Delin, 2002), which can be queried and searched by a normal web browser like any other documents, pictures or videos in World Wide Web (Bröring, et al., 2011; Di, et al., 2010; Zyl, et al., 2009). Originally conceived by Delin, et al. (1999) to describe intelligent wireless sensor networks for coordinated sensing(Delin, 2002), the system later became more and more broadened associating sensors with World Wide Web(Bröring, et al., 2011). This was largely led by Open Geospatial Consortium's (OGC) Sensor Web Enablement (SWE) program started in 2003(OGC, 2012; Bröring, et al., 2011). Main objective of this initiative is to mask glaring disparity among the deployed and future sensor system by defining them in open standards. Taking into account the strong potential for its use in various applications entirely different from the domain of sensor systems actually deployed for, the principle aim is to develop sensor web as a multi-potent and widely usable single window resource for various sensors and its data, much similar to World Wide Web as a source for digital information(Zyl, et al., 2009). For example, by sensor web enablement of earth sensing systems such as weather and air pollution monitoring stations and earth observation satellites the interoperability of data generated by them is taken care by considering all the participating sensor system as single macro-instrument(Delin, 2002). It gives further opportunity to develop applications or raise questions based on system level dynamic information provided by them.

The OGC's SWE specifications provides open standards very similar to that of World Wide Web Consortium (W3C) for maintaining and developing world wide web(W3 2012), to mask the inherent variability in sensors resources. Specifications are essentially focus on the XML based Models and services (Bröring, et al., 2011). The models such as Sensor Model Language (Sensor ML), Observation and Measurement (O& M), and Transducer Model Language (TML) define the capabilities, objectives and limitations of sensors. The services such as Sensor Observation Service (SOS), Sensor Planning Service (SPS), Sensor Alert Service (SAS), Web Notification Service, and Web Processing Service (WPS) extends the capabilities of the sensor resources to search, access, plan, execute and get notifications about observations. It acts as a middle ware for grouping the diverse sensor resources into an interoperable and coordinated entity to develop application over that. Inclusion of details on geo-location in models provides added functionality to extend the applications to geographical realms by integrating with Geographical Information Systems (GIS) and acting as a Spatial Data Infrastructure (SDI) (Torres-Martinez, et al., 2003).

**Sensor webs for Earth System Science:**

Sensor webs provide valuable tool for earth system science by acting as a real time and open ended macro-instrument (Delin, 2002) for earth system observations. It ensures the much needed capabilities to address the dynamism and scalability for earth system science research. The well-timed availability of sensor data supplemented by functionality to search, deploy, plan and task other sensor resources, guarantee the durability and extension of sensor web observations. Standing as an integral part in Internet for live updating, it increases the buffering properties against network collapse by searching and tasking different sensor resources(Zyl et.al., 2009), it also address the need of dynamic observation for various components of earth system. Open ended capability of sensor web, defined by its open standards, models, services and meta-data definition, captures the interaction between discrete sensors. It encapsulates the integration of sensors with their single or prime observation objective of observing atmosphere, hydrosphere, lithosphere or biosphere characteristics to give the opportunity to bring forth holistic view. The opportunity thus the sensor web provides can be considered as defining the earth systems science research by assisting in its multi and inter disciplinary perspective. The vision of earth system science, from the physical basis for understanding earth system through integrative view can be achieved through sensor web enablement of earth observation systems (Torres-Martinez, et al., 2003). For example, the sub regional or local level system understanding of air pollution can be achieved by sensor web enablement and by applications such as air quality models developed over it. The dynamic data requirement for air quality models, such as various location-specific air pollutant emission inventories, meteorological conditions or photochemical phenomenon happening in the atmosphere can be ensured, by sensor web of discrete meteorological stations, air pollution monitoring station and also large global level emissions inventories. The live or near real time data provided by the sensor web along with defining all these data components in set of standards gives opportunity for addressing the dynamic understanding through models and up-scaling it[5] with various other data such as MODIS type satellite observations or physiological data about how the population is effected by air pollution. As hinted above, the application programming interface provided by the sensor web enablement Web Processing Services (WPS) will be taking care of the observation dynamism and scalability issues.

**Sensor webs for social ecological system sustainability:**

To ensure sustainability in multiple interacting "bio-geo-physical" unit and diverse institutions and social actors in social ecological system, system level understanding and inculcation of adaptive characteristics is required (Holling, 2001). For that, new form of doing science rooted in social understanding, adopting new methodologies, using qualitative / quantitative models, harnessing experiences and extracting inverse impact relationship from undesirable consequences to ways for their avoidance is the need of time as observed by Fikert et al. (2002). A realization of the limitations of the current way of doing science in compartmentalized is also required. A synergy between the understanding on the physical, chemical, biological and ecological systems, and of the social systems is to be built up not disregarding experience from other unconventional knowledge systems. "Ecological intelligence" (Daniel, 2009) that "allow us to comprehend systems in all complexity, as well as the interplay between the natural and man-made worlds" and "radical transparency" (Daniel, 2009) that assures universal access to all information is also to be cultivated to ensure sustainability. Sensor webs could be one among the steps to move forward in this line, by establishing tools for multivariate pluralistic monitoring of social ecological system to generate dynamic perception, conflict resolution and for devising appropriate adaptive strategy.

Recent technological and industrial advancement in micro fabrication of electronics especially in sensing technology and wireless communication devices is facilitating reduction of cost for sensor systems and its inclusion in mobile phones. Main achievement of this trend would be wide-reaching of low cost sensors to common public, thus facilitate to ensure participatory approach for pluralistic monitoring of social ecological system and understanding its dynamics. Sensor webs can help in integrating immensely heterogeneous personal sensors and in disseminating location based contextual information. The upcoming projects such as 'Air Quality Egg' (#Air QualityEgg, 2012) are best example of participatory sensor development, deployment and its integrated web based application. This community led, community funded project aims to setup distributed real time air quality sensors and its data sharing networks (Borden 2012). By adopting participatory approach, it strengthens the communities' to develop awareness and empower them for negotiations for better air quality of their immediate surroundings. The project is developed as low cost alternative, minimal expert knowledge and user friendly air quality sensing devices using open source Nanode boards and network of it using Internet of things web application cosm.com (Internet of things web application, 2012). Emphasis is given on spatial resolution, frequency and breadth of air pollution information rather than only accuracy of data. By doing so it intends to develop stand alone sensing devices with low resource and man power intervention to sustain in long run (Borden, 2012). The aim is to learn the pattern and process from the data and to invent the intervention measures which would be more effective in implementations. This kind implementation has greater potential to be applied in Indian scenario, where governance is started to reaching grass root level. The communities can experiment through these systems to know about their air quality. The information openness (Olsson et.al., 2004) provided by sensor web along with governance spreading to the grass root level through Panchayat raj can take effective actions. With required limited expertise, they can uphold and realize the power of self determination to have better living condition, and can impose policy level interventions at local ward or Panchayat level. In brief, using such systems affected communities or villages can gain major advantages. As in the case of villages such as Cuddalore in Tamil Nadu, India (SIPCOT area CEM, 2012), where community monitoring initiatives are currently being carried out using conventional methods, the effectiveness of such endeavors could be further improved. It will further strengthen and widen such endeavors giving much more negotiation power to the community against environmental contamination/hazards or against activities that are not in line with sustainable development and social welfare.

**Conclusion:**

Similar to World Wide Web that has helped achieve information pervasiveness in our day to day life, the sensor web has high potential to achieve better understanding of earth system and to instigate social ecological system sustainability. It will be a great tool for achieving the pressing goals of current and future environmental research.